\documentclass[twocolumn,floatfix]{revtex4-1}%
\usepackage{graphicx}
\usepackage[dvipdfmx]{color}
\usepackage{amsmath,amssymb,amsfonts}
\usepackage{mathrsfs}
\usepackage{mathtools}
\usepackage{here}
\usepackage{color}
\usepackage{bm}
\def\d{{\partial}}

\def\e{{\epsilon}}
\def\k{{ {\bm k} }}

\def\q{{ {\bm q} }}

\def\w{{\omega}}

\allowdisplaybreaks[4]

\begin{document}
\title{
Nematic and chiral superconductivity emerging within the loop-current phase in kagome metals
}
\author{
Rina Tazai$^1$, Youichi Yamakawa$^2$, and Hiroshi Kontani$^2$
}
\date{\today } 

\begin{abstract}
The kagome metals $A$V$_3$Sb$_5$ ($A =$ Cs, Rb, K) host multiple symmetry-breaking phases, including charge-density-wave and loop-current orders, and exhibit highly exotic superconductivity with pronounced nematicity and chirality.
Remarkably, even dilute impurities transform this exotic superconducting state into an isotropic $s$-wave state.
These observations pose a challenge to existing theoretical scenarios.
We show that loop-current order induces nematic chiral $d$-wave superconductivity in kagome metals.
The loop-current-induced orbital magnetization (OM) stabilizes one chiral superconducting channels.
This OM-chirality coupling mechanism is generic and applies to pairing driven by either attractive or repulsive interactions.
Furthermore, coexisting loop-current and bond orders give rise to pronounced nematic chiral superconductivity even for an almost $C_6$-symmetric Fermi surface.
For attractive pairing, dilute impurities suppress the chiral state and restore a conventional $s$-wave phase, as observed experimentally.
The theory further predicts a robust $2\times2$ pair-density modulation. 
This study provides key insights into time-reversal-symmetry-breaking exotic superconductivity in kagome metals.

\end{abstract}

\affiliation{
$^1$Center for Emergent Matter Science, RIKEN, Wako, Saitama 351-0198, Japan. \\
$^2$Department of Physics, Nagoya University, Furo-cho, Nagoya 464-8602, Japan.
}
\sloppy

\maketitle

%%%%%%%%%%%%%%%%%%%%%%%%%%%
\section{Introduction}
%%%%%%%%%%%%%%%%%%%%%%%%%%%
In strongly correlated metals, a variety of exotic quantum phase transitions have been discovered over the years, leading to intriguing symmetry breaking states.
Recently discovered kagome metal $A$V$_3$Sb$_5$ ($A$ = Cs, Rb, K) \cite{Ortiz1,Ortiz2} displays remarkable electronic states arising from a sequence of quantum phases, including the charge bond-order (BO) \cite{BO2,BO3,BO4} and $s$-wave superconductivity with finite impurities \cite{SC1,SC2}.
In addition, $C_6$-symmetry-breaking nematic states emerge in the normal state \cite{BO2,nematicity1,nematicity2,Tazai1} and superconducting (SC) state \cite{Yonezawa-nematicSC,HHWen-nematicSC,Zheng-nematicSC}.
Particularly, time-reversal symmetry (TRS) breaking charge loop-current (LC) state unrelated to spin polarization has garnered significant attention \cite{STM,torque,STM-Yin,STM-Mad,STM-PRB1,STM-PRB2,magnetotropic}.
The LC order parameter involves a ``pure imaginary'' modulation in the hopping integral 
\cite{Varma,Varma2,Affleck,Haldane}.
The microscopic mechanism of the LC order has been investigated in Refs.
\cite{Balents,cLC,BO_theory4,Fernandes}.
The LC is mediated by the BO quantum fluctuations \cite{Tazai2} in kagome metals.

\begin{figure}[htb]
\centering
\includegraphics[width=0.99\linewidth]{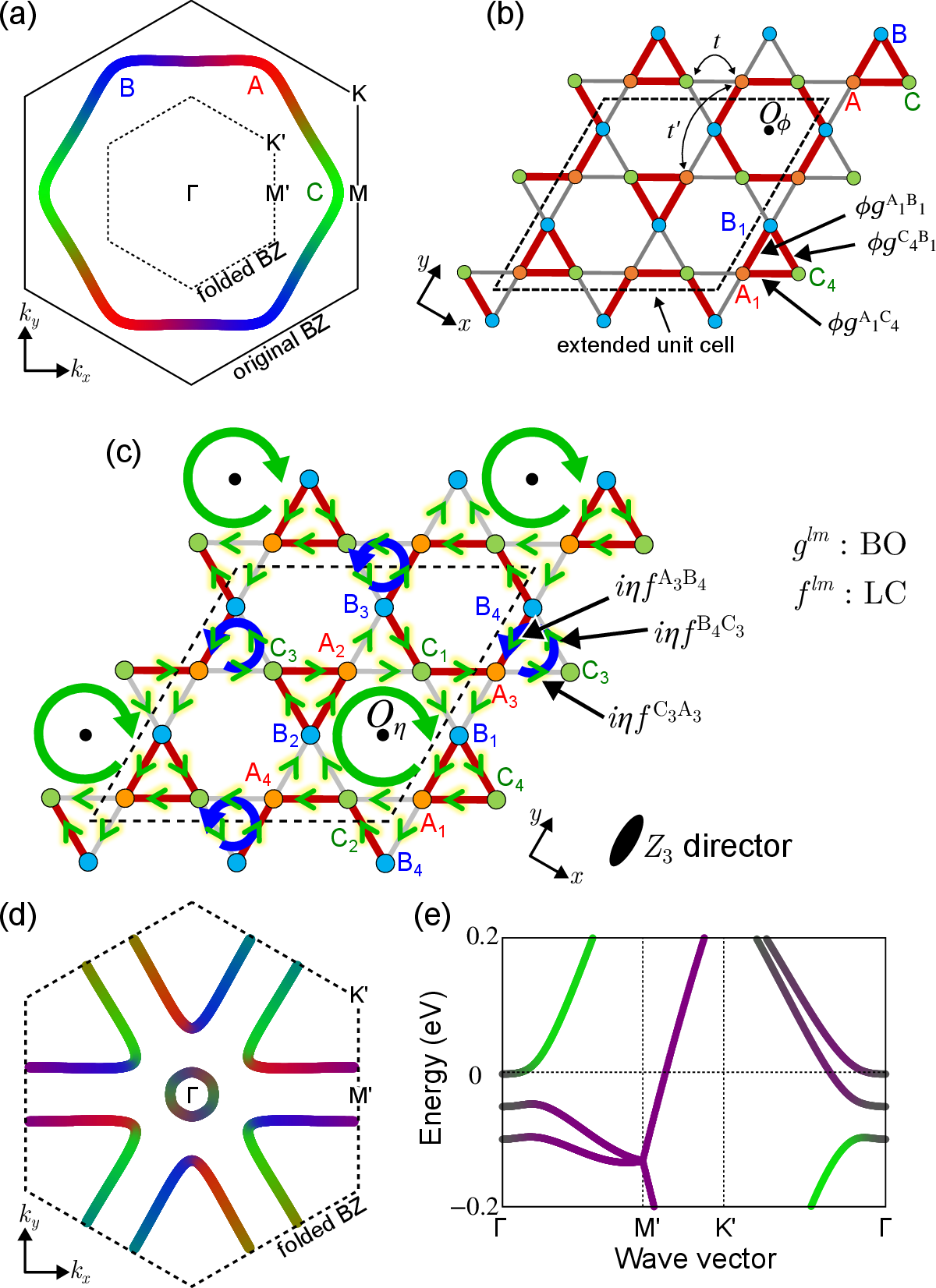}
\caption{
{\bf Kagome lattice model with $2\times2$ BO and LC:} \ 
(a) Fermi surfaces without density-waves in the original BZ.
Red, blue, and green colors represent the weight of the sublattices A, B, and C, respectively.
(b) $2\times2$ TrH BO in the extended unit cell.
%In this paper, we denote site $3\mu-2$ as $A_\mu$, site $3\mu-1$ as $B_\mu$, and site $3\mu$ as $C_\mu$, where $\mu=1,2,3,4$.
$\phi$ is the BO order parameter, and $g^{lm}=g^{ml}$ is the BO form factor.
%$2\boldsymbol{a}_{\mathrm{ab}}$, $2\boldsymbol{a}_{\mathrm{bc}}$ give the two primitive vectors.
(c) $2\times2$ LC in in the extended unit cell.
$i\eta$ is the LC order parameter, and $f^{lm}=-f^{ml}$ is the LC form factor.
The coexistence between the BO in (b) and the LC in (c) leads to the nematic state, whose director is along $y$ (or equivalently $x$) axis. 
The $2\times2$ unit cell is composed of 12 sites denoted as $M_\mu$, where $M$=A,B,C and $\mu=1-4$.
%($M_1,M_2$) and ($M_3,M_4$) form the inversion pairs with respect to $O_\eta$.
(d) Fermi surfaces and (e) band-structure under finite LC ($\eta=0.014$, $\phi=0$) in the folded BZ for $n=11$.
The triple degenerate bands at $E=E_{\rm vHS}$ at $\Gamma$ point are lifted by the LC order.
%$({\bm\eta},{\bm\phi})=({\eta)$ 
}
\label{fig:fig1}
\end{figure}

Figure \ref{fig:fig1} (a) shows the Fermi surface of kagome lattice model denoted in Fig. \ref{fig:fig1} (b).
The $2\times2$ BO and $2\times2$ LC states are schematically shown in Figs. \ref{fig:fig1} (b) and (c), respectively.
In these figures, the BO center ($O_\phi$) and the LC center ($O_\eta$) are out-of-phase \cite{Tazai2,Tazai3}.
Then, the LC + BO coexistence becomes nematic, which is predicted by the Ginzburg-Landau (GL) free energy theory for $T<T_{\rm LC}\ll T_{\rm BO}$ \cite{Tazai2,Tazai3}, and has been experimentally observed \cite{STM-Mad}.

The chiral LC phase accompanies finite uniform orbital magnetization ($M_{\rm orb} \ne 0$), and the LC chirality can be switched by a small out-of-plane magnetic field $B_z$, as evidenced by the anomalous Hall effect (AHE) \cite{AHE1,AHE2,AHE3}, electronic magneto-chiral anisotropy (eMChA) \cite{eMChA,Tazai-eMChA}, and magnetic torque \cite{torque} measurements.
Recent magnetotropic susceptibility measurement reveals a small but heavily anisotropic $M_{\rm orb}$ with tiny hysteresis, consistent with its chiral LC origin \cite{magnetotropic}.
A direct evidence of the chiral LC comes from scanning tunneling microscopy (STM) measurements:
The observed chirality in the quasiparticle interference (QPI) signal is switched by the magnetic field $B_z$ in both $A$=Cs and $A$=K compounds \cite{STM,STM-Yin,STM-Mad,STM-PRB1,STM-PRB2}.
In addition, local staggered magnetic field due to the chiral LC has been reported by $\mu$SR \cite{mSR1,mSR2,mSR3,mSR4} and NMR \cite{NMR} measurements.

Furthermore, striking exotic SC states with nontrivial symmetry breaking appear below $T_{\rm c}$ ($=1\sim3$ K) in kagome metals, as revealed by recent intensive experiments.
For example, STM measurement \cite{STM-Yin} reveals the emergence of (i) TRS breaking chiral SC state that lacks in-plane mirror symmetry. This finding is further supported by the observations of the SC diode effect \cite{SDE-kagome} and giant thermal Hall effect \cite{Yamashita} under zero magnetic field.
In addition, (ii) pronounced nematic anisotropy in the SC gap has been observed \cite{Yonezawa-nematicSC,HHWen-nematicSC,Zheng-nematicSC}.
Furthermore, (iii) prominent modulation in the SC gap, called the pair-density-wave (PDW), has been observed \cite{Roton,STM-Yin} and analyzed theoretically \cite{Ziqiang,STM-Yin}.
Unexpectedly, such exotic pairing states transform into a (iv) conventional $s$-wave state by introducing a small amount of disorder \cite{SC1,SC2,Okazaki-kagome-ARPES}.
At present, the nature of exotic SC states with multiple symmetry breakings in kagome metals --- likely closely linked to the LC order due to the hierarchy $T_{\rm c}\ll T_{\rm LC}$ --- remains a major unresolved problem.

%Such exotic superconducting states with multiple broken symmetries in kagome metals, likely connected to the LC order given $T_{\rm LC} \gg T_{\rm c}$, remain a significant open question.

%At present, such exotic SC states with multiple symmetry breaking in kagome metals, which is expected to be closely related to the LC order because $T_{\rm LC}\gg T_{\rm c}$, remain significant unsolved problem.

%A natural question then arises: What novel SC states and emergent phenomena originate from the LC phase in kagome metals?

In this paper, we propose a simple mechanism of nematic and chiral superconductivity, accompanying the significant PDW component, under the LC phase in the normal state.
Such highly unconventional SC state arises from a simple attractive charge-channel pairing interaction, due to the BO fluctuations and phonons, as suggested by recent experimental and theoretical studies \cite{Tazai1,SC1,SC2,Okazaki-kagome-ARPES}. 
An isotropic chiral $d$-wave SC state is driven by the complex inter-sublattice pair-hopping mechanism under a sole LC order, which becomes relevant at low temperatures.
% when $\eta$ exceeds the threshold $\eta_{\rm c}$, which is on the order of 100K.
When the LC order appears in the BO phase, a prominent nematic chiral SC state is realized due to the mixture of the chiral $d$-wave and $s$-wave states.
%Importantly, $\eta_{\rm c}$ is zero for the nematic chiral SC.  
The present nematic chiral SC state is easily changed to conventional $s$-wave SC by introducing a small amount of impurities, as theoretically discussed in Fe-based SCs \cite{Kontani-PRL2010}.
Furthermore, the present mechanism yields a prominent $2\times2$ PDW component, consistent with experimental observations.
This study provides insights into the fundamental topological nature of SC states arising from the LC phase in kagome metals, such as the SC diode effect and the giant thermal Hall effect under zero magnetic field.

%Interestingly, the pure chiral $d$-wave SC state with topological winding number originates from the sole loop-current order (of order 100K).
% based on the kagome metal models.
%This chiral SC state exhibits pronounced nematicity when the loop-current order and the charge-density-wave coexist.
%In this paper, we investigate the mechanism of superconductivity in the presence of loop-current order, considering the attractive pairing interaction driven by charge-channel fluctuations, as suggested by recent theoretical and experimental studies.

%%%%%%%%%%%%%%%%%%%%%%%%%%%%%%%
%%%%%%%%%%%%%%%%%%%%%%%%%%%%%%%

%At present, the origin and nature of such exotic superconductivity and its close relationship to the normal electronic states have been unsolved.
%These chiral SC states is easily switched to conventional $s$-wave SC by introducing small amount of impurities.
%Interestingly, the chiral SC state with finite winding number would induce interesting topological phenomena.
%, such as giant thermal Hall effect in the SC state.

The microscopic mechanism of the exotic multiple quantum phase transitions has been studied very actively. 
The mean-field theories as well as the renormalization group theories based on the (extended) Hubbard models have been performed in Refs. 
\cite{BO_theory1,BO_theory2,BO_theory3,BO_theory4,Balents,Tazai1,Tazai2,Thomale,Sushkov}.
The present authors discovered that the Star-of-David BO state is driven by the paramagnon-interference \cite{Tazai1}, which is described by the Aslamazov-Larkin vertex corrections that are dropped in the mean-field level approximations
\cite{Onari,fRG,DW_equation,Jianxin,Kontani-rev}.
Note that the BO fluctuations mediate not only attractive SC interaction but also the LC order with breaking TRS
\cite{Tazai1,Tazai2}.

%%%%%%%%%%%%%%%%%
\section{Results}

%%%%%%%%%%%%%%%%%%%%%%%%%%%%%%%%%%%%%%%%%%%%%%%%%%%%%%%%%%%%%%%
\subsection{Kagome lattice model with LC + BO order parameters}

First, we introduce the kagome lattice model:
In the Fermi surface in Fig. \ref{fig:fig1} (a),
each van Hove singularity (vHS) on the original Brillouin zone (BZ) is associated with a single sublattice (A, B, or C), known as sublattice interference \cite{BO_theory1,BO_theory3}.
%Here, each van-Hove singularity (vHS) band-structure is composed of single sublattice (A, B, or C), so called the sublattice interference.
Now, we introduce the BO parameter $\phi$ and the LC one $\eta$:
In Fig. \ref{fig:fig1} (b), the $2\times2$ BO is described as $\delta t_{ij}^{\rm b}=\phi g^{ij}$, where $g^{ij}=g^{ji}=+1$ or $-1$ for the nearest-neighbor sites $(i,j)$ \cite{Tazai3,Nakazawa-QPI,Nakazawa-imp}.
%The Star-of-David BO is realized for $\phi<0$.
Also, the $2\times2$ LC order, described as $\delta t_{ij}^{\rm c}=i\eta f^{ij}$ with $f^{ij}=-f^{ji}=+1$ or $-1$, is depicted in Fig. \ref{fig:fig1} (c)
\cite{Tazai3,Nakazawa-QPI,Nakazawa-imp,Shimura}.
%\rinacom{Definition of $\phi$ and $\eta$ should be added.}
The present LC + BO coexisting state breaks the $C_6$ symmetry because the BO center ($O_\phi$) and the LC center ($O_\eta$) are out-of-phase
\cite{Tazai2,Tazai3}:
see the Methods section A for details.
This $Z_3$ nematic LC + BO coexistence is energetically favorable according to the GL free energy theory for  $T<T_{\rm LC}\ll T_{\rm BO}$ \cite{Tazai2,Tazai3}.

%%%%%%%%%%%%%%%%%%%%%%%%%%%%%%%%%%%%%%%%%%%%%%%%%%%%%%%%%%%%%%%
%\vspace{5mm}
\subsection{SC gap equation for kagome lattice model with $2\times2$ density waves}

In this paper, we set the hopping integrals $t=-0.5$ eV and $t'/t=0.08$ and the electron filling $n=11$ to reproduce the realistic Fermi surface in kagome metals.
Unless otherwise noted, energy is given in units of eV hereafter.
Figures \ref{fig:fig1} (d) and (e) denote the Fermi surfaces and band-structure under finite LC ($\eta=0.014$, $\phi=0$) in the folded BZ, respectively.
Here, single shallow electron-type or hole-type Fermi pocket emerges.
It is caused by the LC-induced hybridization among three vHS points at $\Gamma$ point, which are originally located at M points in the original BZ.
This shallow Fermi pocket has large DOS and exhibits prominent Berry curvature due to the LC order.
We will discuss the significant role of this Fermi pocket in chiral superconductivity.

In kagome metals $A$V$_3$Sb$_5$ ($A$=Cs, Rb, K), an $s$-wave superconducting state without sign reversal is realized in the presence of dense impurities \cite{SC1,SC2,Okazaki-kagome-ARPES}, mediated by an attractive pairing interaction arising from charge-channel fluctuations, including bond-order fluctuations \cite{Tazai1}. Taking this into account, we solve the following gap equation with on-site attractive interaction $g,(>0)$ in the present study:
\begin{eqnarray}
&&\lambda\Delta_m = g\sum_{l}\Gamma_{ml}\Delta_l ,
\label{eqn:gap-eq} \\
&&\Gamma_{ml}= \frac{T}{N}\sum_{\k,n} G^{ml}_{\k}(\e_{n})G^{ml}_{-\k}(-\e_{n}) \Theta(\e_n;\Omega)  ,
\label{eqn:Gamma} 
\end{eqnarray}
where $\lambda$ is the eigenvalue and $\Delta_m$ is the local gap function for sublattice $m \ (=1-12)$.
$\lambda$ reaches unity at $T=T_{\rm c}$. 
${\hat G}_\k(\e_n)$ is the Green function in the presence of the LC ($\eta$) and BO ($\phi$) with Matsubara frequency $\e_n=\pi T(2n+1)$.
$\Theta(\e_n;\Omega)=\Omega^2/((|\e_n|-\pi T)^2+\Omega^2)$ is the cutoff function with $\Omega$; we set $\Omega=0.01$ in the present paper.
Here, we express $\Gamma_{ml}\equiv\Gamma_{ml}^0+\Delta\Gamma_{ml}$: The first term is the real function for $\eta=\phi=0$, and the second term is $\Delta\Gamma_{ml}\equiv\Delta\Gamma_{ml}'+i\Delta\Gamma_{ml}''$, where $\Delta\Gamma'$ ($\Delta\Gamma''$) denote the real (imaginary) parts.
Note that $\Gamma_{mm}\gg|\Gamma_{ml}|$ ($m\ne l$) due to the sublattice interference \cite{BO_theory1,BO_theory3}.
%, and $\Gamma_{ml}$ is approximately independent of $\mu$: $\Gamma_{ml}\approx \Gamma_{ML}$.

%%%%%%%%%%%%%%%%%%%%%%%%%%%%%%%
\begin{figure*}[htb]
\centering
\includegraphics[width=0.7\linewidth]{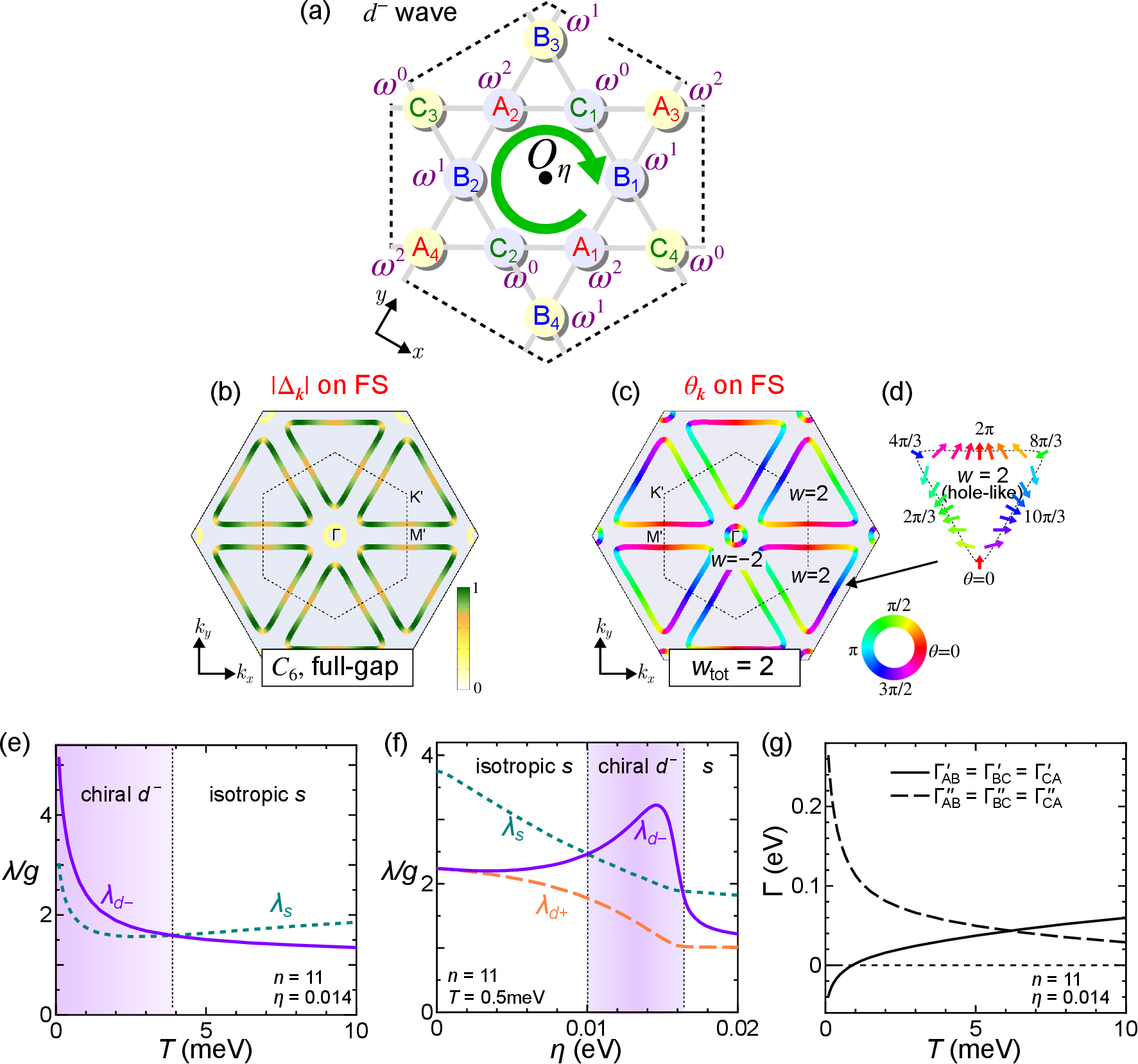}
\caption{
{\bf Chiral $d$-wave superconductivity under a pure LC phase}. \ 
(a) Chiral $d$-wave SC gap ($\chi_d=-1$) under the LC order, where $\w=e^{i 2\pi/3}$.
($s$-wave state corresponds to $\omega=1$.)
%, by setting Im$\Delta_C$=0.
%The number in each site represents the index $\mu$ of the twelve sites $M_\mu$.
%The color of each number represents the sublattice A (red), B (blue), and C (green), respectively.
Here, the LDOS $\rho_{M_\mu}(\e)$ and $|\Delta_{M_\mu}|$ take equal values at $\mu=1$ and $2$, and also at $\mu=3$ and $4$.
(b) $|\Delta_\k|$ and (c) $\theta_\k\equiv \arctan(\Delta_\k''/\Delta_\k')$ on the Fermi surface for chiral $d$-wave state for $\eta=0.014$ at $T=0.1$meV.
$w$ is the winding number of the SC gap for each Fermi pocket.
%Note that $w=\frac1{4\pi}\int {\bm n}(\d_x{\bm n}\times \d_y{\bm n})dk_x dk_y$ for sufficiently small $|\Delta_\k|$, where ${\bm n}$ is the normalized form of $({\rm Re}\Delta_\k,{\rm Im}\Delta_\k,\mu-\e_\k)$.
(d) Vector representation of $\theta_\k$.
Upward arrow corresponds to $\theta=0$, and the size of the arrow is $\sqrt{|\Delta_\k|}$.
(e) Two largest eigenvalues as functions of $T$ for $\eta=0.014$.
The chiral $d$-wave state with $T_{\rm c} <4$meV is realized for $g>0.5$.
%When $g=**$, conventional $s$-wave state is realized at $T_{\rm c}=**$.
%\koncom{The cuttoff energy should be changd to $\Omega=0.01$}.
(f) Three largest eigenvalues as functions of $\eta$ at $T=0.5$meV.
At $\eta=0$, the relation $\lambda_s>\lambda_{d^+}=\lambda_{d^-}$ is satisfied.
$\lambda_{d^-}$ gradually increases for finite $\eta$, and chiral $d^-$-wave state is realized for $\eta=0.01\sim0.016$.
(g) Complex off-diagonal kernel functions $\Gamma_{\rm AB}$, $\Gamma_{\rm BC}$ and $\Gamma_{\rm CA}$ as functions of $T$.
Here, $\rm A\equiv A_3$, $\rm B\equiv B_4$, and $\rm C\equiv C_3$.  
}
\label{fig:fig2}
\end{figure*}
%%%%%%%%%%%%%%%%%%%%%%%%%%%%%%%
%\rinacom{The origin (why this chiral phase in SC gap is selected) is necessary.}
%\rinacom{Phase diagram/ robustness against the parameter is needed.}

%%%%%%%%%%%%%%%%%
%\vspace{5mm}
\subsection{Chiral $d$-wave superconductivity under a pure LC phase}

Hereafter, we express the twelve sublattices as $M_\mu$, where $M$=A, B, or C and $\mu=1\sim4$; as denoted in Fig. \ref{fig:fig2} (a): 
$\mu=1$ and $2$ denote the six sites surrounding the LC center $O_\eta$.
That is, $\mu=1,2 \ (3,4)$ denote the sites participating in the circulating current on a hexagonal (triangular) plaquette.
Then, the SC gap function is expressed as $\Delta_{M_\mu}$, and we introduce the notation ${\bm\Delta}_\mu\equiv(\Delta_{\rm A_\mu},\Delta_{\rm B_\mu},\Delta_{\rm C_\mu})$.
The obtained ${\bm\Delta}\mu$ is nearly independent of $\mu$ in the $s$-wave state.
In contrast, in the $d$-wave state, ${\bm\Delta}\mu$ exhibits a clear $\mu$-dependence (up to approximately 5 \%), corresponding to the $2 \times 2$ PDW component; see Section B of the Methods for details.

First, we analyze the gap equation for $\phi=0$ (without BO).
When $\eta$ is zero, a plane $s$-wave SC state is obtained as the largest eigenvalue.
The second-largest ($\chi_d=\pm1$) eigenvalue corresponds to the double-degenerated chiral $d$-wave solutions.
This double degeneracy is lifted when $\eta$ is finite.
In this paper, we reveal that a single chiral SC state ($\chi_d=+1$ or $-1$) can take over the $s$-wave SC state in the LC phase, because $\Gamma_{ml}$ acquires an effective Aharanov-Bohm phase factor originating from the LC order $\eta\ne0$.
The obtained chiral $d$-wave state ${\bm\Delta}_\mu\propto(1,\w^2,\w)$, which corresponds to $\chi_d=-1$, is depicted in Fig. \ref{fig:fig2} (a) in real space.  
The chirality $\chi_d$ is reversed by changing $\eta$ to $-\eta$.
The gap function on the $b$-th Fermi momentum $\k$ is given as $\Delta_{b,\k}=\sum_m w^{m,b}_{\k}\Delta_m$, where $\w^{m,b}_\k$ is the weight of $m$-th sublattice on the $b$-th band.
Hereafter, we omit the index $b$, as it is uniquely determined for a given Fermi momentum $\k$.

In the chiral $d$-wave state, $\Delta_\k \equiv \Delta_\k' + i\Delta_\k''$ is a complex-valued function of $\k$.
Figures \ref{fig:fig2} (b) and (c) show the obtained $|\Delta_\k|$ and $\theta_\k\equiv \arctan(\Delta_\k''/\Delta_\k')$ on the Fermi surface for the chiral $d$-wave state, respectively, for $\eta=0.014$ at $T=0.1$meV.
Here, $|\Delta_\k|$ exhibits a sizable dependence on $\k$ due to the phase difference of $\Delta_m$.
The vector representation of $\theta_\k$ for a single Fermi pocket is given in Fig. \ref{fig:fig2} (d).
The upward arrow corresponds to $\theta=0$, and the size of the arrow is $\sqrt{|\Delta_\k|}$.
We also denote the winding number of the SC gap for each Fermi pocket, which is defined as $w=\frac1{4\pi}\int {\bm n}(\d_{k_x}{\bm n}\times \d_{k_y}{\bm n})dk_x dk_y$ for sufficiently small $|\Delta_\k|$.
Here, ${\bm n}=(\Delta_\k',\Delta_\k'',\mu-\e_\k)/\sqrt{|\Delta_\k|^2+(\mu-\e_\k)^2}$.

Figure \ref{fig:fig2} (e) shows the largest and second-largest eigenvalues as functions of $T$ for $\eta=0.014$.
Notably, the obtained $\lambda_d$ drastically increases for $T<5$meV.
Therefore, a chiral $d$-wave state with a transition temperature $T_{\rm c} \ll 4$ meV is realized for $g < 0.5$. (Note that $\lambda_x$ in the higher $T$ region ($\gtrsim 0.01$) is overestimated, as the pairing constant $g$ due to electron correlations should decrease with increasing $T$.) Figure~\ref{fig:fig2}(f) shows the three largest eigenvalues as functions of $\eta$ at $T = 0.5$ meV. At $\eta = 0$, the relation $\lambda_s > \lambda_{d+} = \lambda_{d-}$ holds. As $\lambda_{d-}$ gradually increases with $\eta$, the chiral $d^-$-wave state emerges for $\eta$ in the range 0.01-0.016. Thus, the $T_{\rm c}$ of the chiral $d$-wave state is enhanced by the LC order.

In Section B of the Methods, we explain the emergence of the $2 \times 2$ PDW component induced by the background LC + BO phase. 
In the chiral $d$-wave superconducting state, the PDW component becomes $\sim 5$\%, which is comparable to experimental reports \cite{STM-Yin}. In contrast, the PDW component is $\ll 1$\% in the $s$-wave state, as given by the second-largest eigenvalue of the SC gap equation.
The PDW due to $2\times2$ density wave has been analyzed based on the GL theory in Ref. \cite{Fisher-GL}, while we derive the PDW state directly by solving the gap equation for the twelve-site unit cell model.

%%%%%%%%%%%%%%%%%
%\vspace{5mm}
\subsection{Inter-sublattice complex pair-hopping mechanism in the LC phase}

Here, we discuss why chiral $d$-wave SC is realized in the LC phase due to the charge-channel attractive pairing interaction.
The LC-induced band-hybridization gap suppresses the diagonal kernel function $\Gamma_{mm}\sim N(0)\log(\Omega/T)$, leading to the reduction of the eigenvalue $\lambda$.
(Here, $N(\e)$ is the density-of-states.)
However, $\Gamma_{ml}$ with $m\ne l$ becomes complex functions for $\eta\ne0$, and its magnitude is drastically magnified at low temperatures, as shown in Fig. \ref{fig:fig2} (e).
Then, the phase factor $\theta_{ml}=\arctan(\Gamma_{ml}''/\Gamma_{ml}')$ prefers the phase difference between $\Delta_m$ and $\Delta_l$, working as the inter-sublattice "complex Josephson coupling".
%More detailed explanation is given in Appendix **. \koncom{Appendix:}
For this reason, the chiral $d$-wave SC state emerges in the LC phase.

To verify this intuitive understanding,
%To understand why chiral $d$-wave SC state is causaed by $\Delta\Gamma_{lm}'(<0)$ and $\Delta\Gamma_{lm}''\ne0$,
we analyze a simplified gap equation for three original sublattices A (=A$_3$), B (=B$_4$), and C (=C$_3$), by neglecting the $\mu$-dependence of $\Delta_{M_\mu}$:
By referring to Fig. \ref{fig:fig2} (g),
we introduce the following simple complex kernel functions:
\begin{eqnarray}
g\hat{\Gamma}=
\begin{pmatrix}
   \lambda_0 & a-ib &a+ib \\
   a+ib & \lambda_0 & a-ib \\
   a-ib & a+ib & \lambda_0 
\end{pmatrix}
\end{eqnarray}
where $a$ and $b$ are real constants.
The eigenvaues and the gap functions are given as
\begin{eqnarray}
&&\cdot \lambda=\lambda_0+2a : \ \bm{\Delta}\propto(1,1,1)  =  {\rm isotropic \ s}
\\
&&\cdot \lambda=\lambda_0-a-\sqrt{3}b : \ \bm{\Delta}\propto(1,\w^2,\w)  =  {\rm chiral \ d^-}
\\
&&\cdot \lambda=\lambda_0-a+\sqrt{3}b : \ \bm{\Delta}\propto(1,\w,\w^2)  =  {\rm chiral \ d^+}
\end{eqnarray}
Thus, we find the following results:
(i) Isotropic $s$-wave SC gap is realized when $\eta=0$ because $b=0$ and $a>0$.
(ii) Chiral $d^+$ or $d^-$ SC state is realized for $|b|>\sqrt{3}a$, depending on the sign of $b$.
(iii) The sign of $b$ depends on $n$ even for a fixed $\bm{\eta}$.
(iv) The ratio $|b/a|$ increases as $T$ is lowered; see below.
%as we explained based on the diagrammatic method. 

In the LC state, the LDOS pattern shown in Fig. \ref{fig:fig2}(a) belongs to the $D_{6h}$ point group, so the chiral $d$-wave state ($E_{2g}$) and the $s$-wave state ($A_{1g}$) belong to the different irreducible representation.
In addition, TRS breaking in the LC order lifts the degeneracy between $\lambda_{d+}$ and $\lambda_{d-}$.

%%%%%%%%%%%%%%%%%%%%%%%%%%%%%%%
\begin{figure}[htb]
\centering
\includegraphics[width=0.99\linewidth]{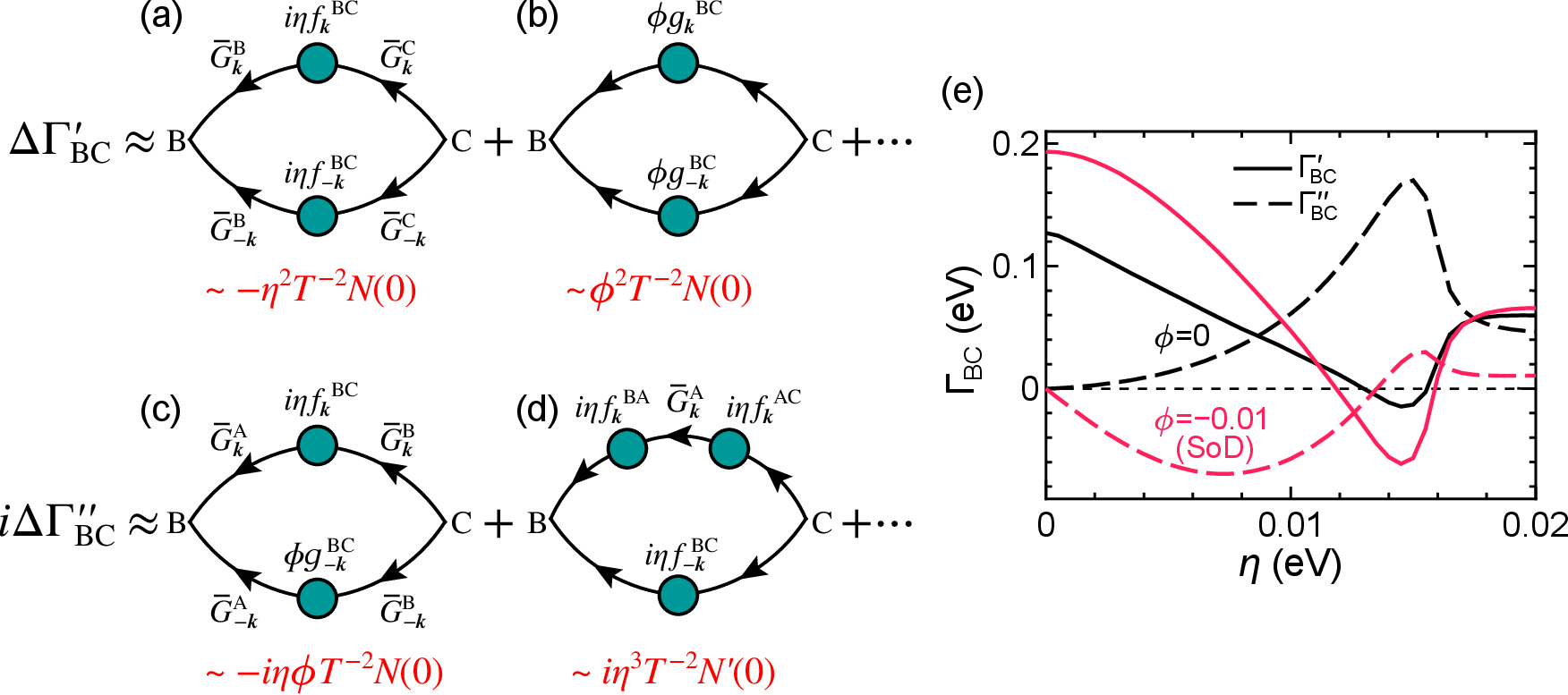}
\caption{
{\bf TRS breaking pair-hopping mechanism given by off-diagonal kernel functions}: \ 
Leading-order expansion terms of (a)(b) $\Delta\Gamma_{\rm BC}'$ and (c)(d) $i\Delta\Gamma_{\rm BC}''$ with respect to $\eta,\phi$. 
${\bar G}^{M}_{k}$ is the diagonal Green function for $\eta=\phi=0$, and $N(0)$ is the density-of-states.
Here, $M$ = A, B, C belong to an original unit-cell.
(e) Numerical results of $\Delta\Gamma_{\rm BC}'$ and $\Delta\Gamma_{\rm BC}''$ as functions of $\eta$, at $T=0.5$ meV.
For both $\phi=0$ and $0.01$, the obtained results for $\eta\ll0.01$ are well explained by the processes (a)-(d).
}
\label{fig:diagram}
\end{figure}
%%%%%%%%%%%%%%%%%%%%%%%%%%%%%%%

Here, we present an analytic explanation for emergence of the prominent $\Delta\Gamma_{\rm BC}'$ and $\Delta\Gamma_{\rm BC}''$ at low temperatures in the LC phase.
In kagome metals, the $12\times12$ Green function for $\eta=\phi=0$ is approximately sublattice-diagonal around the vHS points:
${\bar G}^{ml}(k) \approx {\bar G}^{m}(k)\cdot\delta_{m,l}$ \cite{Shimura}.
Based on this simplification,
the leading-order expansion terms of $\Delta\Gamma_{\rm BC}$ with respect to $\eta,\phi$ are expressed in Figs. \ref{fig:diagram} (a)-(d).
%Based on this simplification, the $\eta^2$-term for $\Delta\Gamma_{\rm BC}'$ is given as Fig. \ref{fig:diagram} (a).
In the holded BZ, the vHS point is $\k={\bm0}$, where the relations $g^{lm}=\pm1$ ($g^{ml}=g^{lm}$) and $f^{lm}=\pm1$ ($f^{ml}=-f^{lm}$) are satisfied for the nearest $(m,l)$ pair. 
In the case of $\phi=0$, 
$\Delta\Gamma_{\rm BC}'\propto -\eta^2T^{-2}N(0)$ and 
$i\Delta\Gamma_{\rm BC}''\propto i\eta^3T^{-2}N'(0)$, which are given by
Figs. \ref{fig:diagram} (a) and (d), respectively.
Importantly, this mechanism becomes relevant at low temperatures, providing a natural explanation for the numerical results given in Fig. \ref{fig:fig2} (e). 
In the case of $\phi\ne0$, 
large imaginary part $\eta\phi$-linear term $i\Delta\Gamma_{\rm BC}''\propto -i\eta\phi T^{-2}N(0)$ is caused by Fig. \ref{fig:diagram} (c). 
Therefore, the prominent chiral SC state originates from the coexisting LC + BO state even once $\eta\phi$ is finite, as we will explain in the following.

Figure \ref{fig:diagram} (e) presents the numerical results of $\Delta\Gamma_{\rm BC}'$ and $\Delta\Gamma_{\rm BC}''$ as functions of $\eta$, at $T=0.5$ meV.
For both $\phi=0$ and $0.01$, the obtained prominent $\eta$-depencences for $\eta\ll0.01$ are well explained by the processes in Figs. \ref{fig:diagram} (a)-(d).

In the Methods section C, we discuss the eigenvalues ($\lambda_d$, $\lambda_s$) as functions of $n$.
The obtained $\lambda_d$ takes sizable values for $n\sim11.0$, where the LC-induced small Fermi pocket appears around the $\Gamma$ point.
In this case, the LC-induced complex kernel function $\Gamma_{ml}$ ($m\ne l$) given by Fig. \ref{fig:diagram} takes a large value because $|{\bar G}_\k^A ||{\bar G}_\k^B|$ is large at $\k={\bm0}$.
In contrast, $\lambda_s$ is nearly independent of $n$.

%%%%%%%%%%%%%%%%%%%%%%%%%%%%%%%
\begin{figure*}[htb]
\centering
\includegraphics[width=0.7\linewidth]{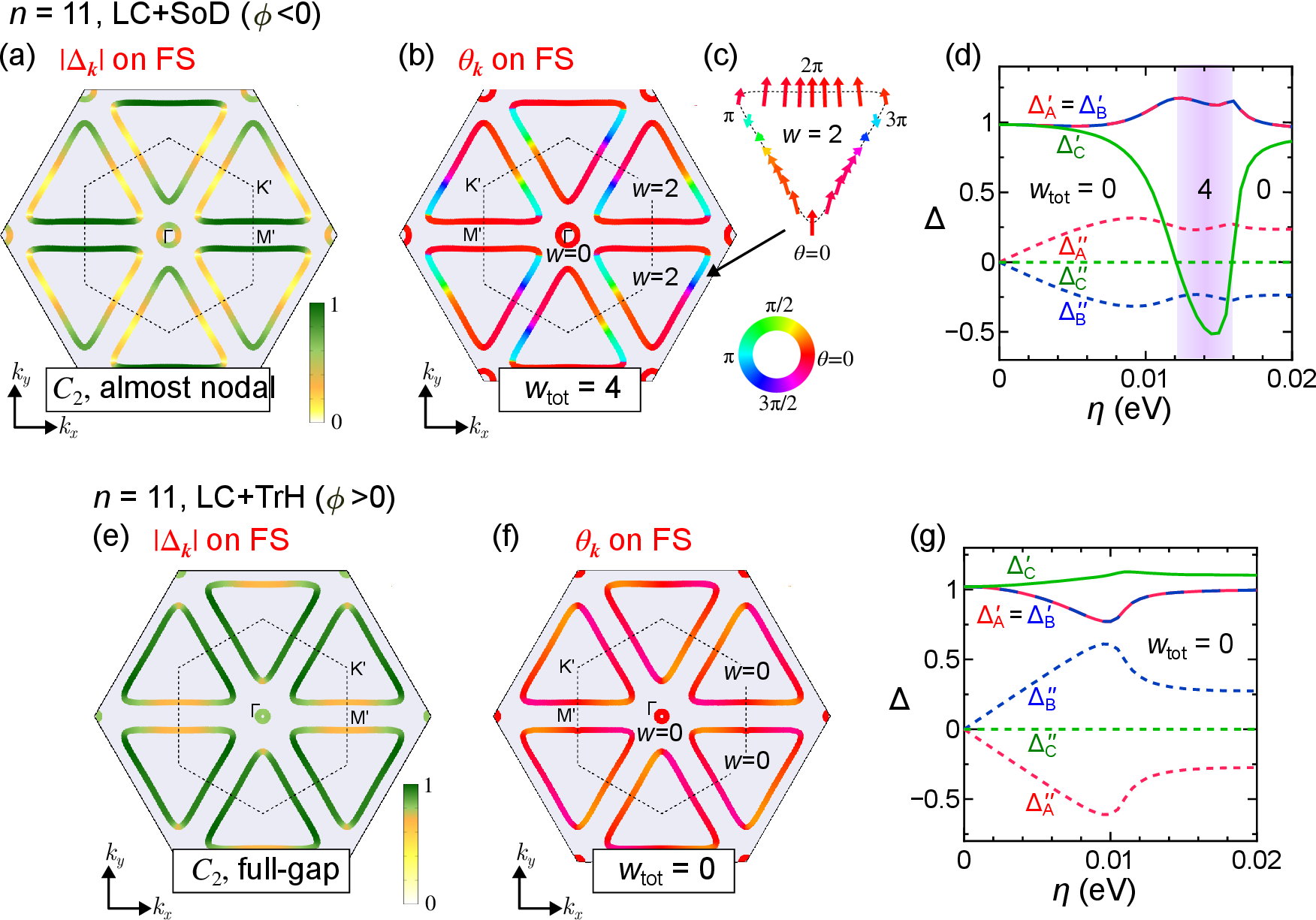}
\caption{
{\bf Nematic chiral superconductivity driven by LC under (a-d) SoD BO ($\phi=-0.01$) and (e-h) TrH BO ($\phi=0.01$)}: \ 
(a)(b) Obtained $|\Delta_\k|$ and $\theta_\k$ for $\phi=-0.01$ (SoD) and $\eta=0.014$. 
The gap structure exhibits pronounced nematic anisotropy along the AB direction (the $k_y$-axis), despite the nearly $C_6$-symmetric Fermi surface.
(c) Vector representation of $\theta_\k$.
Upward arrow corresponds to $\theta=0$, and the size of the arrow is $\sqrt{|\Delta_\k|}$.
(d) Obtained $\Delta_{\rm A}$, $\Delta_{\rm B}$, and $\Delta_{\rm C}$ as functions of $\eta$ under $\phi=-0.01$.
Here, $\rm A\equiv A_3$, $\rm B\equiv B_4$, and $\rm C\equiv C_3$.
(e)(f) Obtained $|\Delta_\k|$ and $\theta_\k$ for $\phi=0.01$ (TrH) and $\eta=0.01$. 
The gap structure also exhibits sizable nematic anisotropy.
%(g) Vector representation of $\theta_\k$.
(g) Obtained $\Delta_{\rm A}$, $\Delta_{\rm B}$, and $\Delta_{\rm C}$ as functions of $\eta$ under $\phi=0.01$.
}
\label{fig:nematicSC}
\end{figure*}
%%%%%%%%%%%%%%%%%%%%%%%%%%%%%%%

%%%%%%%%%%%%%%%%%
%\vspace{5mm}
\subsection{Nematic chiral SC states driven by LC + BO order}

In kagome metals,
the chiral LC order appears inside the star-of-David BO.
The nematic LC + BO coexisting state breaks the $C_6$ symmetry when the chiral LC order and the BO are out-of-phase ({\it i.e.}, $O_\phi\ne O_\eta$).
This $Z_3$ nematic LC + BO coexistence is energetically favored based on the GL free energy theory for $T<T_{\rm LC}\ll T_{\rm BO}$ \cite{Tazai2,Tazai3}, and it is observed experimentally \cite{STM-Mad}.
In the following, we analyze the gap equation in the nematic LC + BO coexisting state.
Because the LDOS pattern shown in Fig. \ref{fig:Z3} (c) belongs to $D_{2h}$ point group, the chiral $d$-wave state ($|l_z|=2$) belongs to the $A_{1g}$ representation.
Therefore, the $s$ + chiral $d$-wave state is expected to be realized, with sizable nematicity and chirality.
We first discuss the SoD BO state with $\phi=-0.01$:
Figures \ref{fig:nematicSC} (a) and (b) exhibits the obtained $|\Delta_\k|$ and $\theta_\k$, respectively, for $(\eta,\phi)=(0.014,-0.01)$.
The gap structure exhibits pronounced nematic anisotropy along the AB direction (the $k_y$-axis), despite the fact that the Femri surface is nearly $C_6$ except for the small Femri pocket around $\Gamma$-point.
A vector representation of $\theta_\k$ is presented in Fig. \ref{fig:nematicSC} (c).
Upward arrow corresponds to $\theta=0$, and the size of the arrow is $\sqrt{|\Delta_\k|}$.
Figure \ref{fig:nematicSC} (d) exhibits the obtained $\Delta_{\rm A}$, $\Delta_{\rm B}$, and $\Delta_{\rm C}$ as functions of $\eta$ under $\phi=-0.01$.
Here, $\rm A\equiv A_3$, $\rm B\equiv B_4$, and $\rm C\equiv C_3$.
We also discuss the TrH BO state with $\phi=0.01$:
Figures \ref{fig:nematicSC} (e) and (f) exhibits $|\Delta_\k|$ and $\theta_\k$, respectively, for $(\eta,\phi)=(0.01,0.01)$.
The gap structure also exhibits sizable nematic anisotropy.
%A vector representation of $\theta_\k$ is shown in Fig. \ref{fig:nematicSC} (g).
Figure \ref{fig:nematicSC} (g) exhibits the obtained $\Delta_{\rm A}$, $\Delta_{\rm B}$, and $\Delta_{\rm C}$ as functions of $\eta$ under $\phi=0.01$.

Thus, in both SoD BO and TrH BO cases, the $s$ + chiral $d$-wave state with prominent nematicity and chirality is realized.
%Therefore, the chiral SC state is naturally expected to be realized in real kagome metlas.
Note that the Anderson theorem indicates that the chiral SC state is fragile against impurities and randomness.
Therefore, the LC-induced chiral SC state will be changed to simple isotropic $s$-wave state by introducing small amount of impurities.

Experimentally, sizable nematic anisotropy in the SC gap has been observed in Refs. \cite{Yonezawa-nematicSC,HHWen-nematicSC,Zheng-nematicSC}, irrespective of the nearly $C_6$ symmetry Fermi surfaces in the CDW phase.
The present nematic chiral SC state will give a natural explanation for the nematic SC gap in kagome metals.
%The present nematic chiral SC state will not be fragile against V-site impurities because the gap function is local.
However, the isotropic $s$-wave SC state is realized by introducing impurities.

In the Methods section C, we present some supporting numerical results for the nematic chiral SC state in the LC + BO phase.
Note that the LC-induced chiral SC state also appears for $n\sim10$, where one or two small Fermi pockets appear around $\Gamma$ point; see the Methods section D.

%Finally, we study the quantum transport phenomena characteristics of the LC-induced chiral SC state.

%%%%%%%%%%%%%%%%%%%%%%
%\vspace{5mm}
\section{Summary}

Rich spontaneous symmetry-breaking phenomena in metals are central to condensed matter physics. In kagome metals $A$V$_3$Sb$_5$ ($A$=Cs, Rb, K), the normal electronic states exhibit the chiral loop-current without inversion symmetry, nematicity with breaking rotational symmetry, and chirality without in-plane mirror symmetry. 
The superconducting states in kagome metals also exhibit (i) chirality without time-reversal symmetry, (ii) prominent nematicity, in addition to the (iii) $2\times2$ pair-density-wave state. Unexpectedly, such an exotic pairing states changes to a (iv) conventional $s$-wave state due to dilute impurities. The understanding of this unconventional superconductivity have hardly been achieved to this day.

In this study, we proposed a mechanism of nematic and chiral $d$-wave superconductivity, with significant PDW component, under the coexistence of the LC and BO in the normal state.
Such highly exotic SC states in kagome metals arise from a simple attractive charge-channel pairing interaction, due to the BO fluctuations and phonons.
Under a sole loop-current order, an isotropic chiral $d$-wave SC state is driven by the complex inter-sublattice pair-hopping mechanism.
Importantly, this mechanism becomes relevant at low temperatures.
%When only the LC order emerges in the normal state, pure chiral $d$-wave SC state is realized.
When LC order coexists with the BO, nematic chiral SC states due to $s$ + chiral $d$ state is realized.  
The obtained $|\Delta_\k|$ in Fig. \ref{fig:nematicSC} exhibits sizable nematic anisotropy, in spite of the fact that the Fermi surfaces possess almost perfect $C_6$ symmetry.
Therefore, the above-mentioned (i)-(iii) are naturally explained.
In the present mechanism, such unconventional pairing state is easily changed to conventional $s$-wave SC by impurities, consistently with (iv) mentioned above.
Furthermore, the present mechanism yields a prominent $2\times2$ PDW component, consistent with experimental observations.
The present study provides insights into the topological nature of exotic SC states arising from the LC phase in kagome metals.

%aaaaa
%Notably, the present chiral SC state with finite winding number would induce interesting topological phenomena, such as giant thermal Hall effect in the SC state.
%This study provides insights into the fundamental nature of exotic superconducting states arising from the LC phase in kagome metals.

%%%%%%%%%%%%%%%%%%%%%%%%%%%%%%%
\begin{figure}[htb]
\centering
\includegraphics[width=0.99\linewidth]{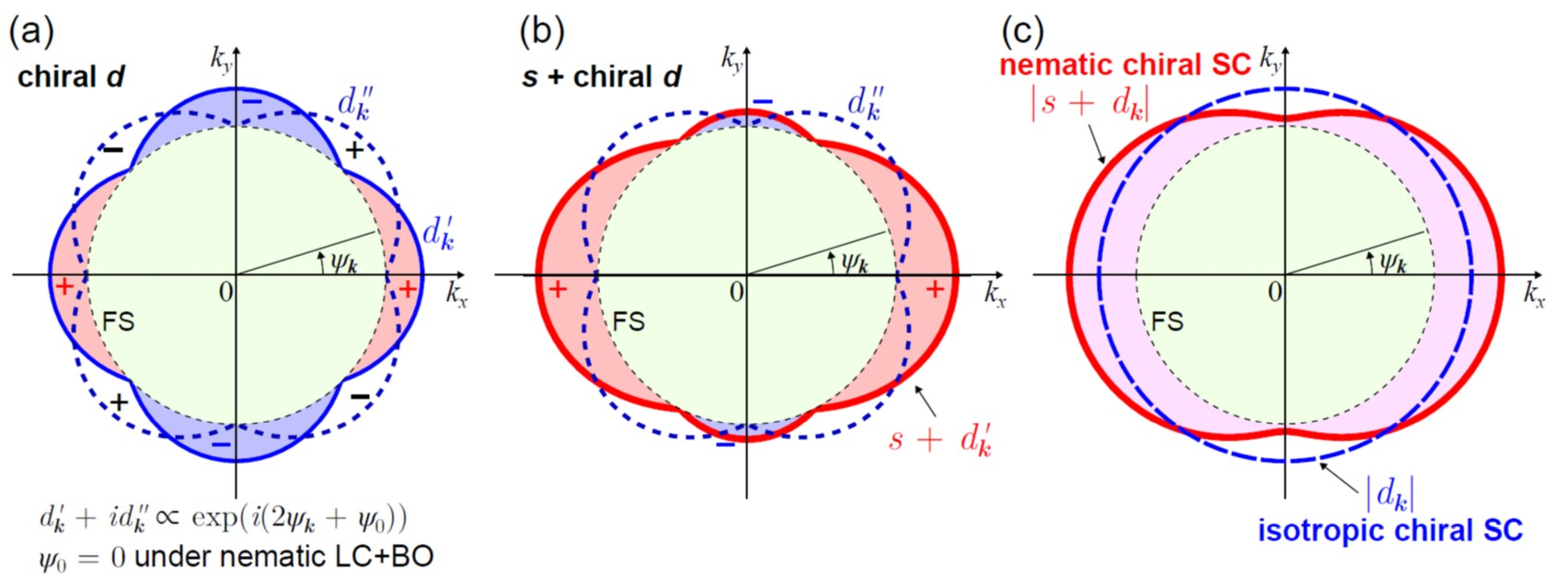}
\caption{
{\bf Nematic chiral SC state realized by chiral $d$ + $s$ state}: \ 
(a) Chiral $d$-wave gap ($d_\k=d_\k'+id_\k''$) on a circular Fermi surface.
Here, $d_k=d_0\exp(2i\psi_\k+i\psi_0)$ with $\psi_\k=\arctan(k_y/k_x)$ for $\psi_0=0$.
%The dashed circle represents the $E = 0$ level, and $\psi_\k=\arctan(k_y/k_x)$.
(b) $s$ + chiral $d$-wave gap, given by introducing the isotropic $s$-wave gap ($s$=real) into the chiral $d$-wave gap.
The constant phase $\psi_0$ is fixed to $0$, $2\pi/3$, or $4\pi/3$ under $Z_3$ LC + BO nematic state \cite{Tazai3}. 
Note that $\Delta_\k$ is chiral if $|d_0|>|s|$.
(c) Absolute value of the SC gap $\Delta_\k = d_\k + s$, which exhibits remarkable nematicity for $|d_0|\sim |s|$. %\rinacom{chical wo chiral}
}
\label{fig:schematic}
\end{figure}
%%%%%%%%%%%%%%%%%%%%%%%%%%%%%%%%%

Here, we present an intuitive explanation for the nematic chiral SC state due to $s$ + chiral $d$ SC state obtained by the present study.
Figure \ref{fig:schematic} (a) represents the chiral $d$-wave gap ($d_\k=d_\k'+id_\k''$) and isotropic $s$-wave gap ($s$=real) on a circular Fermi surface.
The dashed circle represents the $E = 0$ level, and $\psi_\k=\arctan(k_y/k_x)$.
Here, $d_k'=d_0\cos(2\psi_\k+\psi_0)$ and $d_k''=d_0\sin(2\psi_\k+\psi_0)$,
where $d_0$ is a real constant, and the constant phase $\psi_0$ is fixed to $0$, $2\pi/3$, or $4\pi/3$ under $Z_3$ LC + BO nematic state \cite{Tazai3}.
Figure \ref{fig:schematic} (c) exhibits the absolute value of the SC gap in the $s$ + chiral $d$ SC state, $\Delta_\k = d_\k + s$, which exhibits remarkable nematicity for $|d_0|\sim |s|$.
Its director rotates with $\psi_0$.
Notably, the SC gap $\Delta_\k$ is chiral if $|d_0|>|s|$.

%Notably, the present chiral SC state with finite winding number would induce interesting topological phenomena, such as giant thermal Hall effect in the SC state.
This study provides insights into the fundamental nature of topological superconducting states arising from the LC phase in kagome metals.
For example, recently observed giant thermal Hall effect in CsV$_3$Sb$_5$ below $T_{\rm c}$ \cite{Yamashita} would be driven by from present TRS-breaking chiral SC state.
Also, the present study would represent an important step toward understanding the chirality in the superconducting state in kagome metals observed in Ref. \cite{STM-Yin}.
The significance of $p$-orbital electrons was pointed out in Ref. \cite{STM-Yin}.
In addition, the origin of the SC diode effect under zero magnetic field \cite{SDE-kagome} and $(4/3)\times(4/3)$ PDW state \cite{Roton,Ziqiang} is of particular interest.

%\koncom{Future scope: transport such as the THE, SC diode, PDW $(4/3)x(4/3)$, etc}

%\koncom{Comments on previous theoretical studies, such as GL analysis by Mark Fisher.}

%\appendix

%%%%%%%%%%%%%%%%%%%%%%%%%%%%%%%%%%%%%%%%%%%%%%%%%%%%%%%%%%%%%%%%%%
\newpage

%\makeatletter
%\renewcommand{\thefigure}{\arabic{figure}}
%\renewcommand{\theequation}{S\arabic{equation}}
%\renewcommand{\figurename}{Extended Data Fig.}
%%
%\renewcommand{\thefigure}{\arabic{figure}}
%\renewcommand{\theequation}{\arabic{equation}}
%\renewcommand{\figurename}{Supplementary Fig.}
%\makeatother
%\setcounter{figure}{0}
%\setcounter{equation}{0}
%\setcounter{page}{1}
%\setcounter{section}{1}

%\begin{widetext}
%\begin{center}
%{\bf \large 
%[Supplementary Information] 
%\\ \vspace{3mm}
%{\large
%}}
%\end{center}

%\begin{center}
%Rina Tazai$^{1}$, Youichi Yamakawa$^{2}$, and Hiroshi Kontani$^2$
%\end{center}

%\begin{center}
%\textit{
%$^1$Yukawa Institute for Theoretical Physics, Kyoto University, Kyoto 606-8502, Japan \\
%$^2$Department of Physics, Nagoya University, Nagoya 464-8602, Japan \\
%}
%\end{center}

%\end{widetext}
%%%%%%%%%%%%%%%%%%%%%%%%%%%%%%%%%%%%%%%%%%

\section{Methods} 

%%%%%%%%%%%%%%%%%%%%%%%%%%%%%
\subsection{A: Z$_3$ nematic LC + BO coexisting state}

\begin{figure}[htb]
\centering
\includegraphics[width=0.99\linewidth]{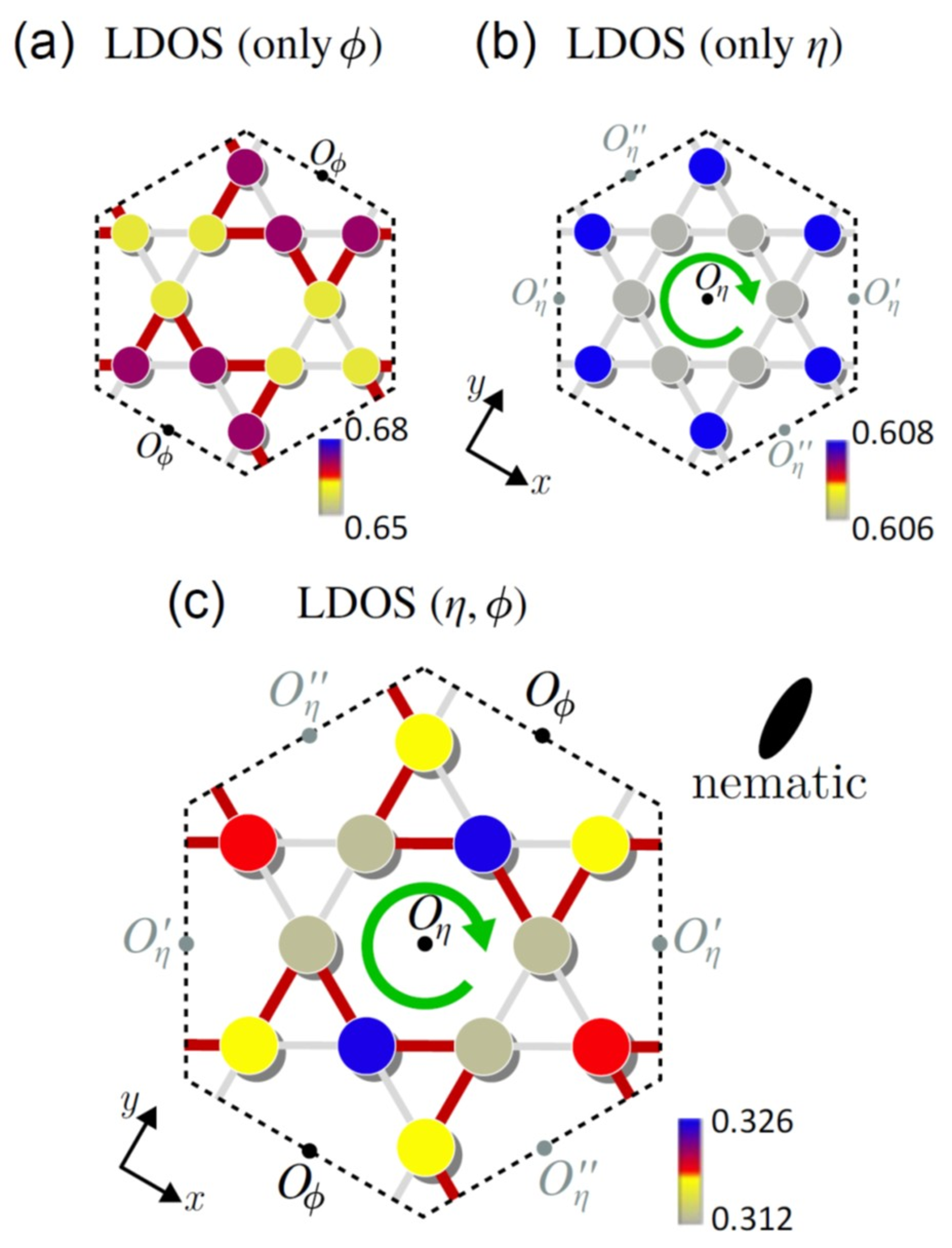}
\caption{
{\bf $Z_3$ nematic LC + BO coexisting state}: \ 
(a) LDOS in the BO phase illustrated in Fig. \ref{fig:fig1} (b), where the triple-$\q$ order parameter is ${\bm\phi}\propto(1,1,1)$.
The center of $C_6$ symmetry is shown as $O_\phi$.
(b) LDOS in the LC phase illustrated in Fig. \ref{fig:fig1} (c), where the triple-$\q$ order parameter is ${\bm\eta}\propto(1,1,1)$.
The center of $C_6$ symmetry is shown as $O_\eta$.
(c) LDOS in the $Z_3$ nematic LC + BO coexisting phase, where $C_6$ symmetry center is absent. 
The director is rotated by $-2\pi/3$ ($2\pi/3$) by shifting the LC center from $O_\eta$ to $O_\eta'$ ($O_\eta''$), which corresponds to ${\bm\eta}'\propto (1,-1,1)$ (${\bm\eta}''\propto (1,1,-1)$).
}
\label{fig:Z3}
\end{figure}

In this paper, we analyzed the SC gap equation based on the kagome lattice model with $2\times2$ density waves.
A $2\times2$ density wave is described by the combination of three density waves with the wavevector $\q=\q_1, \ \q_2, \ \q_3$, called the triple-$\q$ order.
Here, we express the set of triple-$\q$ BO orders as ${\bm\phi}\equiv(\phi_1,\phi_2,\phi_3)$, where $\phi_n$ is the order parameters at $\q=\q_n$.
Then, $\phi_1$, $\phi_2$, and $\phi_3$ give the BO modulation between the nearest ab-, bc- and ca-bonds, respectively.
Figure \ref{fig:fig1} (b) represents the BO for ${\bm\phi}=\phi(1,1,1)$, which corresponds to $\phi_1=\phi_2=\phi_3$.
Then, the center of $C_6$ symmetry is $O_\phi$.
This BO induces the $2\times2$ modulation of the local density-of-states (LDOS), as shown in Fig. \ref{fig:Z3} (a).

In the same way, the $2\times2$ LC is given by the triple-$\q$ order parameters ${\bm\eta}\equiv(\eta_1,\eta_2,\eta_3)$.
$\eta_1$, $\eta_2$, and $\eta_3$ give the LC orders that induce the charge current along ab-, bc- and ca-bonds, respectively.
Then, the chirality of the triple-$\q$ LC state is described as $\chi_{\rm LC}\equiv{\rm sgn}\{\eta_1\eta_2\eta_3\}$. 
The uniform orbital magnetization $M_{\rm orb}$ is proportional to $\chi_{\rm LC}$.
Figure \ref{fig:fig1} (c) represents the LC for ${\bm\eta}=\eta(1,1,1)$, where the center of $C_6$ symmetry is $O_\eta$.
The induced $2\times2$ modulation of the LDOS is shown in Fig. \ref{fig:Z3} (b).
As explained in Refs. \cite{Tazai2,Tazai3}, the center of $C_6$ symmetry moves to $O_\eta'$ ($O_\eta''$) for ${\bm\eta}'=\eta(-1,-1,1)$ (${\bm\eta}''=\eta(-1,1,-1)$), without changing $\chi_{\rm LC}$.

As explained in the main text, the LC order parameter attaches the effective Aharanov-Bohm phase to conduction electrons.
For this reason, the chiral $d$-wave SC state,
$(\Delta_{\rm A},\Delta_{\rm B},\Delta_{\rm C})\propto(\w,\w^2,1)$ or $(\w^2,\w,1)$,
can be realized under the $2\times2$ LC state.
Its chirality depends on the LC chirality as well as the electron filling.

%%%%%%%%%%%%%%%%%%%%%%%%%%%%%%%
%%%%%%%%%%%%%%%%%%%%%%%%%%%%%%%

Next, we consider the coexistence between the BO and the LC.
When $O_\phi$ and $O_\eta$ are different, the $C_6$ symmetry center disappears, leading to the electronic nematic state.
Figure \ref{fig:Z3} (c) represents the obtained nematic LDOS pattern under ${\bm\phi}=\phi(1,1,1)$ and ${\bm\eta}=\eta(1,1,1)$.
The director is rotated by $-2\pi/3$ ($+2\pi/3$) by shifting ${\bm\eta}$ to ${\bm\eta}'$ (${\bm\eta}''$).
In this $Z_3$ nematic state, the chiral $d$-wave SC state and $s$-wave SC coexists as a consequence of reduced symmetry.
This fact leads to the nematic chiral SC state as revealed in this paper.
The nematicity and the chirality in the SC state becomes prominent for $n\sim11$, where the original Fermi surface is close to three vHS points.
Under the triple-$\q$ order, three vHS points form a small Fermi pocket as depicted in Fig. \ref{fig:fig1} (d), due to the LC-induced band reconstruction.
This reconstructed Fermi pocket gives rise to sizable complex number kernel function $\Gamma_{ml}$ ($m\ne l$) shown in Fig. \ref{fig:diagram}, which is the origin of the prominent nematic and chiral SC state in the gap equation.

%%%%%%%%%%%%%%%%%%%%%%%%%%%%%
\subsection{B: $2\times2$ pair-density-wave (PDW) state}

Recent experimental study \cite{STM-Yin} reports the emergence of the $2\times2$ pair-density-wave (PDW) state in $A$V$_3$Sb$_5$ below $T_{\rm c}$.
The modulation of the period $2a$ SC gap due to the PDW is about 5 \%.
Such large modulation cannot be simply explained because the modulation of the local DOS due to LC + BO order is less than 1\%.
Here, we explain that the (nematic) chiral $d$-wave state driven by the LC order exhibits sizable $2\times2$ PDW modulation.

\begin{figure}[htb] 
\centering
\includegraphics[width=0.99\linewidth]{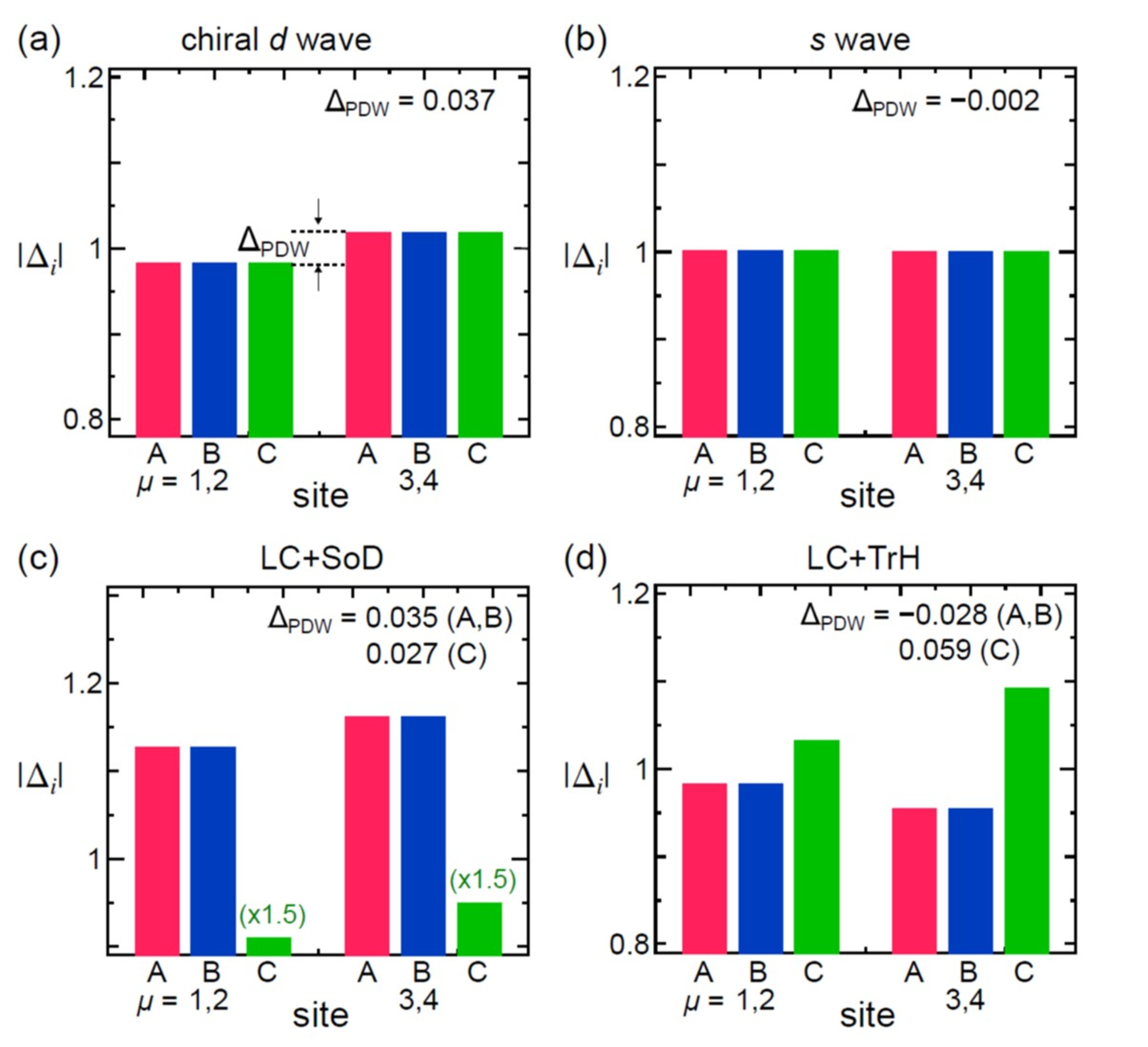}
\caption{
{\bf $2\times2$ PDW in the LC and LC + BO phases: $|\Delta_{M_1}|$ and $|\Delta_{M_3}|$ ($M$=A,B,C)}: \
(a) $|\Delta_{M_\mu}|$ in the chiral $d$-wave state for $(\eta,\phi)=(0.014,0)$ at $T=0.1$ meV.
Finite $\Delta_{\rm PDW}\equiv |\Delta_{M_3}|-|\Delta_{M_1}|$ gives the $2\times2$ PDW component.
$\Delta_{\rm PDW}$ increases to $0.08$ for $\eta=0.016$. 
(b) $|\Delta_{M_\mu}|$ in the $s$-wave state, which is the second largest eigenvalue in the numerical study for (a).
Here, the PDW component is very small.
(c) $|\Delta_{M_\mu}|$ in the nematic chiral SC state in the LC + BO (SoD) state with $(\eta,\phi)=(0.014,-0.01)$ at $T=0.1$ eV.
(d) those in the LC + BO (TrH) state with $(\eta,\phi)=(0.01,0.01)$ at $T=0.1$ eV.
}
\label{fig:PDW}
\end{figure}

First, we discuss the PDW under the pure LC state without BO.
In this case, $|\Delta_{M_\mu}|$ is independent of $M$=A, B, and C. 
Figure \ref{fig:PDW} (a) exhibits the $|\Delta_{M_\mu}|$ in the chiral $d$-wave state for $(\eta,\phi)=(0.014,0)$ at $T=0.1$ meV.
(Both $|\Delta_\k|$ and $\theta_\k$ in the chiral $d$-wave state for the same parameter are shown in Fig. \ref{fig:fig2} in the main text.)
In this case, $|\Delta_{M_1}|=|\Delta_{M_2}|$ deviates from $|\Delta_{M_3}|=|\Delta_{M_4}|$.
Importantly, finite $\Delta_{\rm PDW}\equiv|\Delta_{M_3}|-|\Delta_{M_1}|$ means the $2\times2$ PDW modulation, due to the translational symmetry breaking in the SC state.
%In this case, the ratio $\Delta_{\rm PDW}/\Delta$ reaches $\sim10$ \%.
Note that $\Delta_{\rm PDW}$ increases to $0.08$ for $\eta=0.016$. 
The second largest eigenvalue corresponds to the $s$-wave state, whose $|\Delta_{M_\mu}|$ is shown in Fig. \ref{fig:PDW} (b).
In this case, the difference $\Delta_{\rm PDW}$ is very tiny and invisible.

Next, we discuss the PDW under the $Z_3$-nematic LC + BO states discussed in the main text.
Figures \ref{fig:PDW} (c) and (d) exhibit the $|\Delta_{M_\mu}|$ in the $s$ + chiral $d$-wave state for $(\eta,\phi)=(0.014,-0.01)$ and $(\eta,\phi)=(0.01,0.01)$, respectively. 
Note that $\phi<0$ ($\phi>0$) represents the SoD (TrH) BO.
(Both $|\Delta_\k|$ and $\theta_\k$ for the same parameter are shown in Fig. \ref{fig:nematicSC} in the main text.)
%In both SC states, the $2\times2$ PDW modulation $\Delta_{\rm PDW}$ is $\sim5$ \%.
Notably, the obtained $s$ + chiral $d$-wave SC state exhibits sizable nematicity expressed by $|\Delta_{\rm A_\mu}|-|\Delta_{\rm C_\mu}|$.
To summarize, the present $s$ + chiral $d$-wave SC state driven by the $Z_3$ nematic LC + BO exhibits prominent chirality and nematicity, both of which are widely observed in kagome metals.

%%%%%%%%%%%%%%%%%%%%%%%%%%%%%%%
%%%%%%%%%%%%%%%%%%%%%%%%%%%%%%%

%%%%%%%%%%%%%%%%%%%%%%%%%%%%%
\subsection{C: $n$-dependence of $\lambda_x$ ($x=s,d$)}

In the main text, we performed the numerical study only for the electron filling $n=11$.
Here, we verify the chiral $d$-wave SC state is realized for finite range of $n$.

\begin{figure}[htb]
\centering
\includegraphics[width=0.9\linewidth]{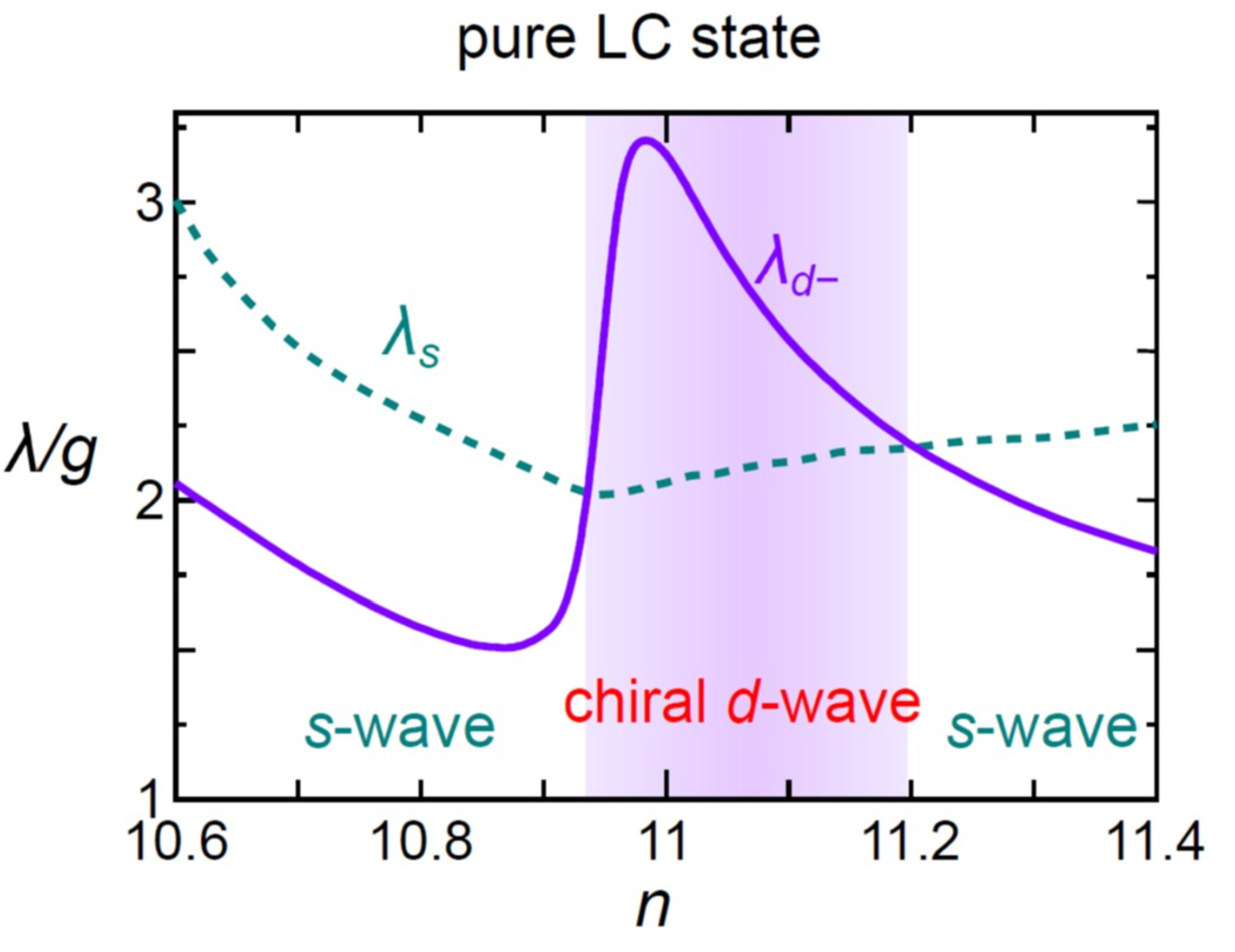}
\caption{
{\bf $n$-dependence of $\lambda_x$ ($x=s,d$)}: \ 
The largest and second-largest eigenvalues as functions of $n$, where $(\eta,\phi)=(0.014,0)$ at $T=0.5$meV.
The chiral $d$-wave state dominates over the $s$-wave state in the region $10.9\lesssim n \lesssim11.2$, where single small Fermi pockets appears around $\Gamma$ point.
The region of chiral $d$-wave state is enlarges for larger $\eta$.
}
\label{fig:n-dependence}
\end{figure}

Figure \ref{fig:n-dependence} represents the first and second SC eigenvalues as functions of $n$, where $(\eta,\phi)=(0.014,0)$ at $T=0.5$meV.
The chiral $d$-wave state dominates over the $s$-wave state in the region $10.9\lesssim n \lesssim11.2$, where single small Fermi pockets appears around $\Gamma$ point.
As we discuss in the main text, the chiral $d$-wave SC state emerges when three vHSs composed of all sublattices (A, B, C) form a small hybridized Fermi surface around the $\Gamma$ point under the triple-$\q$ LC state.
Importantly, $\lambda_d$ is drastically enlarged for $10.9\lesssim n \lesssim11.2$.

%%%%%%%%%%%%%%%%%%%%%%%%%%%%%%%
%%%%%%%%%%%%%%%%%%%%%%%%%%%%%%%

%\koncom{To check the robustness and to clarify the realization condition:}

%%%%%%%%%%%%%%%%%%%%%%%%%%%%%
\subsection{D: Supporting numerical results}

\begin{figure}[htb]
\centering
\includegraphics[width=0.99\linewidth]{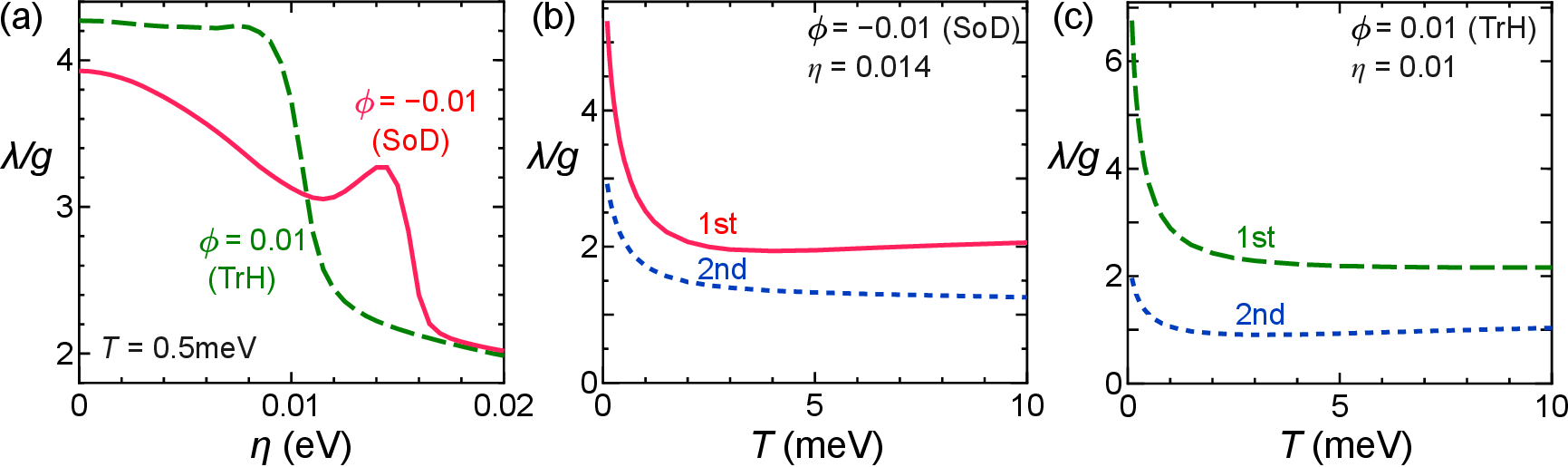}
\caption{
{\bf Eigenvalue $\lambda$ in the nematic LC + BO state}: \ 
(a) $\lambda$ as a function of $\eta$ under the BO: $\phi=-0.01$ (SoD) and $\phi=0.01$ (TrH).
(b) Two largest eigenvalues as a function of $T$ under $(\eta,\phi)=(0.014,-0.01$).
(c) Two largest eigenvalues under $(\eta,\phi)=(0.01,0.01$).
}
\label{fig:LCplusBO}
\end{figure}

In the present paper, we proposed the emergence of the nematic chiral SC state under the LC + BO phase.
Here, we present more detailed numerical results to confirm the validity of the present study.
Figure \ref{fig:LCplusBO} (a) shows the obtained $\lambda$ as a function of $\eta$ under the BO: $\phi=-0.01$ (SoD) and $\phi=0.01$ (TrH).
Figure \ref{fig:LCplusBO} (b) exhibits two largest eigenvalues as a function of $T$ under $(\eta,\phi)=(0.014,-0.01$).
In addition, two largest eigenvalues under $(\eta,\phi)=(0.01,0.01$) are given in Fig. \ref{fig:LCplusBO} (c).

%%%%%%%%%%%%%%%%%%%%%%%%%%%%%%%
%%%%%%%%%%%%%%%%%%%%%%%%%%%%%%%

\begin{figure}[htb]
\centering
\includegraphics[width=0.99\linewidth]{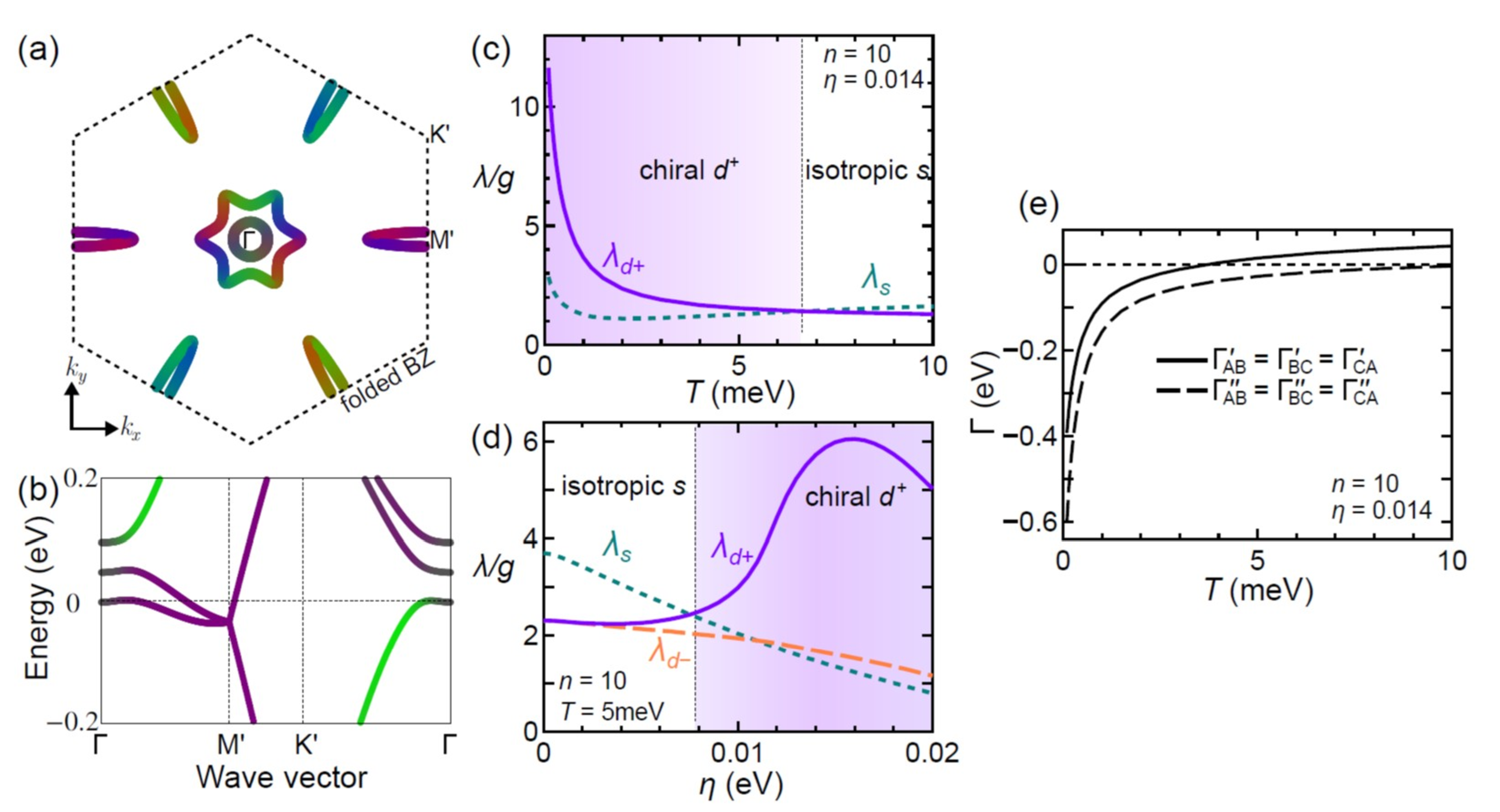}
\caption{
{\bf LC-induced chiral $d$-wave state at $n=10.0$}: \ 
(a) Fermi surface and (b) band-structure for $(\eta,\phi)=(0.01,0)$ at $n=10$.
(c) The largest and second-largest eigenvalues as functions of $T$.
The chiral $d$-wave state with the transition temperature $T_{\rm c} <6.5$meV is realized for $g>0.6$.
(d) The three largest eigenvalues as functions of $\eta$ at $T=0.5$meV.
At $\eta=0$, the relation $\lambda_s>\lambda_{d^+}=\lambda_{d^-}$ is satisfied.
$\lambda_{d^+}$ gradually increases for finite $\eta$, and chiral $d^+$-wave state is realized for $\eta>0.008$.
(e) Complex off-diagonal kernel functions $\Gamma_{\rm AB}$, $\Gamma_{\rm BC}$ and $\Gamma_{\rm CA}$ as functions of $T$.
Here, $\rm A\equiv A_3$, $\rm B\equiv B_4$, and $\rm C\equiv C_3$.  
}
\label{fig:n10}
\end{figure}

In the present paper, we studied the chiral $d$-wave state under the LC phase in kagome metals, by focusing the electron filling $n\sim11$.
To verify the robustness of the present analysis,
we perform the numerical study for $n=10$.
Figure \ref{fig:n10} (a) represents the Fermi surface and (b) band-structure for $n=10$ and $(\eta,\phi)=(0.01,0)$, where hole-like Fermi pockets appear.
Figure \ref{fig:n10} (c) shows the obtained largest and second-largest eigenvalues as functions of $T$.
The chiral $d$-wave state with the transition temperature $T_{\rm c} <6.5$meV is realized for $g>0.6$.
Figure \ref{fig:n10} (d) represents the three largest eigenvalues as functions of $\eta$ at $T=0.5$meV.
At $\eta=0$, the relation $\lambda_s>\lambda_{d^+}=\lambda_{d^-}$ is satisfied.
$\lambda_{d^-}$ gradually increases for finite $\eta$, and chiral $d^-$-wave state is realized for $\eta>0.008$.
The complex off-diagonal kernel functions $\Gamma_{\rm BC}$, $\Gamma_{\rm BC}$ and $\Gamma_{\rm CA}$, which are the driving force of the chiral SC state, are shown in Figure \ref{fig:n10} (e).
Here, $\rm A\equiv A_3$, $\rm B\equiv B_4$, and $\rm C\equiv C_3$.  
Thus, the LC-driven chiral $d$-wave SC state is expected to be realized for both $n\sim10$ and $n\sim11$.

%%%%%%%%%%%%%%%%%%%%%%%%%%%%%%%
%%%%%%%%%%%%%%%%%%%%%%%%%%%%%%%

%%%%%%%%%%%%%%%
%\vspace{5mm}
%{\bf Numerical analysis}: \
%%%%%%%%%%%%%%
%In this study, we derived the $12\times12$ kernel functions in the SC gap equaiton, $\Gamma_{ml}$, by performing the $\k$-summation and Matsuraba-frequency summation in Eq. (\ref{eqn:Gamma}) numerically. The Brillouin zone is divided into $1024^2$ $\k$ meshes. The number of $\k$ meshes is fine enough to achieve reliable numerical accuracy (within a few persent) for the present model parameters ($T\sim0.5$meV, $\eta,\phi\sim0.01$, $\Omega=10$meV).

%%%%%%%%%%%%%%%
%\vspace{5mm}
%{\bf Data availability}: \  
%%%%%%%%%%%%%%%
%The data generated by the present numerical study have been deposited in the Figshare database. 

%\subsection{Acknowledgments}
\acknowledgements
We are grateful to T. Matsushita for enlightening discussions on the chiral $d$-wave superconductivity.
We are also grateful to M. Yamashita, Y. Matsuda, T. Shibauchi, S. Suetsugu, and S. Onari for fruitful discussions.

%{\bf General}: We are grateful to for useful discussions.

%%%%%%%%%%%%%%%%%%%%%%%%%%%%%%%%%%%%%%%%

%%%%%%%%%%%%%%%%%%%%%%%%%%%%%%%%%%%%

%%%%%%%%%%%%%%%%%%
%\vspace{5mm}
%{\bf Funding}: \
%This study has been supported by Grants-in-Aid for Scientific Research from MEXT of Japan (JP24K00568, JP24K06938, JP22K14003), and by ``Correlated Design Science'' (JP25H01246 and JP25H01248) KAKENHI on Transformative Research Areas from JSPS of Japan.

%%%%%%%%%%%%%%%%%%
%\vspace{5mm}
%{\bf Author contributions}: \
%R.T. and Y.Y. executed the calculations in discussion with H.K., and all authors contributed to writing the paper. 

%%%%%%%%%%%%%%%%%%
%\vspace{5mm}
%{\bf Competing interests}: \
%The authors declare that they have no competing interests.

\end{document}